\documentclass[traditabstract]{aa}

\usepackage{amsmath}
\usepackage{amssymb}
\usepackage[varg]{txfonts}
\usepackage{epsfig,graphicx}
\usepackage[colorlinks=true, linkcolor=blue, citecolor=blue, urlcolor=blue]{hyperref}
\usepackage{natbib}
\bibpunct{(}{)}{;}{a}{}{,} 
\newcommand{\txe}{{\text{e}}}

\newcommand{\bfx}{{\boldsymbol{x}}}

\begin{document}

\title{Composite biasing in Monte Carlo radiative transfer}

\titlerunning{Composite biasing}

\author{%
Maarten Baes\inst{\ref{UGent}} 
\and
Karl D. Gordon\inst{\ref{STScI},\ref{UGent}}
\and
Tuomas Lunttila\inst{\ref{Onsala}}
\and
Simone Bianchi\inst{\ref{INAF}}
\and
Peter Camps\inst{\ref{UGent}}
\and
Mika Juvela\inst{\ref{Helsinki}}
\and
Rolf Kuiper\inst{\ref{Tubingen}} 
}

\institute{%
Sterrenkundig Observatorium, Universiteit Gent, Krijgslaan 281 S9, B-9000 Gent, Belgium\label{UGent}
\and
Space Telescope Science Institute, 3700 San Martin Drive, Baltimore, MD 21218, USA\label{STScI}
\and
Chalmers University of Technology, Department of Earth and Space Sciences, Onsala Space Observatory, 439 92, Onsala, Sweden\label{Onsala}
\and
INAF -- Osservatorio Astrofisico di Arcetri, Largo E. Fermi 5, 50125, Florence, Italy\label{INAF}
\and
Department of Physics, PO Box 64, University of Helsinki, 00014, Helsinki, Finland\label{Helsinki}
\and
Institute of Astronomy and Astrophysics, University of T\"ubingen, Auf der Morgenstelle 10, D-72076 T\"ubingen, Germany\label{Tubingen}
}

\abstract{Biasing or importance sampling is a powerful technique in Monte Carlo radiative transfer, and can be applied in different forms to increase the accuracy and efficiency of simulations. One of the drawbacks of the use of biasing is the potential introduction of large weight factors. We discuss a general strategy, composite biasing, to suppress the appearance of large weight factors. We use this composite biasing approach for two different problems faced by current state-of-the-art Monte Carlo radiative transfer codes: the generation of photon packages from multiple components, and the penetration of radiation through high optical depth barriers. In both cases, the implementation of the relevant algorithms is trivial and does not interfere with any other optimisation techniques. Through simple test models, we demonstrate the general applicability, accuracy and efficiency of the composite biasing approach. In particular, for the penetration of high optical depths, the gain in efficiency is spectacular for the specific problems that we consider: in simulations with composite path length stretching, high accuracy results are obtained even for simulations with modest numbers of photon packages, while simulations without biasing cannot reach convergence, even with a huge number of photon packages.  
}


\maketitle

\section{Introduction}

The Monte Carlo (MC) simulation method can be defined as a rather direct transcription of natural stochastic processes into computing terms, and is a widely applied method to simulate complex problems. MC simulations are routinely used for transport problems of all kinds of particles, including neutrons, cosmic rays, neutrinos and photons. In the case of photon transport, or radiative transfer, the MC method has grown to be the most popular method, both for line radiation and continuum radiation \citep[for overviews, see e.g.,][]{Dupree2002, 2011BASI...39..101W, 2013ARA&A..51...63S, Marchuk2013}.

All modern Monte Carlo radiative transfer (MCRT) methods incorporate a number of acceleration methods to speed up the simple MC loop. The essential ingredient of all acceleration methods is the assignment of a weight to each photon package, which can alter during its life cycle. The weight of a photon package is equivalent to the fraction of the original luminosity carried by that photon package, or the number of photons contained within it. 

Several of the most popular acceleration methods in MCRT are based on the technique of biasing. Biasing consists of generating random numbers from a different probability density function (pdf) from the physical one, and correcting for this biased probability by adapting the weight of the photon package. This technique can be applied in order to sample particular parts of the pdf more heavily than others. 

In the context of MCRT, biasing is used in several optimisation techniques. Probably the most widespread application of biasing is the technique of forced scattering \citep{Cashwell1959, 1970A&A.....9...53M, 1977ApJS...35....1W}, which is built into most modern radiative transfer codes in some form. Another direct application of biasing is the principle of biased emission, by which the location from which and/or the propagation direction into which photon packages are emitted are biased \citep{1984ApJ...278..186Y, 2005A&A...440..531J}. An advanced application of biasing is the case of polychromatism, in which photon packages containing photons of different wavelengths are used in the simulation \citep{2006MNRAS.372....2J}.

While biasing is undeniably a powerful technique to make MCRT simulations more efficient, it also comes with its potential dangers. One aspect that should be kept in mind is that if one uses biasing to sample particular parts of the pdf more heavily than one should naturally do, other parts of the pdf are automatically sampled less heavily. In other words, while biasing is designed to decrease the noise where desired, the logical price to pay is that the noise will increase in other places. This implies that biasing needs to be used with care, and this can limit the applicability of the technique in "blind" applications where little a priori information about the system is known. 

A second important aspect is that biasing adds an additional weight factor to each photon package. Small weight factors are usually not a problem, but photon packages with large weight factors can be a significant source of noise, and this can completely ruin the potential advantages of biasing \citep[see e.g.,][]{2005A&A...440..531J, 2006MNRAS.372....2J}. 

In this paper, we further explore the use of biasing in MCRT simulations. In Section~{\ref{CompositeBiasing.sec}} we go deeper into the idea of biasing and we introduce a specific approach of biasing (which we call composite biasing) that automatically avoids the problem of large weight factors. In the following two sections we present two novel applications of (composite) biasing in MCRT. In Section~{\ref{MultipleComponents.sec}} we focus on the emission of photon packages from a source that is composed of multiple components. In Section~{\ref{PathlengthStretching.sec}} we focus on the problem of penetrating through an optically thick medium. We demonstrate there that the use of composite biasing to the generation of random optical depths can be an elegant way to deal with this problem. A discussion and conclusions are presented in Section~{\ref{Conclusions.sec}}.

\section{Composite biasing}
\label{CompositeBiasing.sec}

The biasing principle is used in many MC applications and can be a very effective acceleration technique \citep{Dupree2002}. It is known as importance sampling in MC numerical integration and computer graphics \citep{1978JCoPh..27..192L, Kalos2009}, and as umbrella sampling in computational molecular sciences \citep{1977JCoPh..23..187T, Kastner2011}.

In MCRT, biasing is typically used in the following manner. Assume that, in the course of the life cycle of a photon package, we need to determine a random event $x$ from a given pdf $p(x)$.\footnote{This event can be, for example, the location of the emission, the path distance that can be covered before the next interaction, or the propagation direction after a scattering event. The probability density function $p(x)$ is completely determined by the radiative transfer problem that is being considered.}  The most obvious way to proceed is to generate a random $x$ directly from the pdf $p(x)$ using any of the techniques for univariate or multivariate nonuniform random number generation available in the literature \citep[see e.g.,][]{Devroye1986, Hormann2004}. Sometimes, however, we want to promote certain parts of the domain more in order to increase the signal-to-noise ratio there. For example, one could have good reasons to prefer more photon packages to be emitted in a certain direction, even though the emission is intrinsically isotropic \citep{1984ApJ...278..186Y, 2005A&A...440..531J}. The technique of biasing consists of generating $x$ not from the original pdf $p(x)$, but from a different pdf $q(x)$. The biased behaviour is corrected for by altering the weight of the photon package with a weight factor 
\begin{equation}
w(x) = \frac{p(x)}{q(x)}
\label{wx}
\end{equation}
In MC integration, the pdf $q(x)$ is often a scaled or translated version of the original pdf, but this is not required: the biased pdf can in principle be chosen completely arbitrarily. The only restriction is that $q(x)>0$ for all $x$ where $p(x)>0$, except for those $x$ that one knows will not contribute to the final tally.

As already mentioned, one of the limitations of the applicability of the biasing technique is the possible appearance of large weight factors. Since both $p(x)$ and $q(x)$ are normalised probability functions, it is unavoidable that the weight function $w(x)$ will be larger than unity in some part of the domain. If the weight of a photon package is boosted significantly, it can become a serious source of noise. The choice of the biased pdf is therefore a delicate job. On the one hand, $q(x)$ should be chosen such that it does promote those parts of the domain that are desired. On the other hand, it should be such that, when combined with the original pdf, the weight factor is never boosted to excessively large values.

A general way to combine these two requirements is what we propose as composite biasing. Assume that we have an intrinsic pdf $p(x)$ (for example, a uniform distribution on the unit sphere to represent isotropic emission), and that we would ideally prefer a biased pdf $q(x)$ (for example, a distribution strongly peaked towards one specific direction). If $p(x)$ and $q(x)$ are radically different, it is possible (and even likely) that the weight factor (\ref{wx}) becomes very large in some part of the domain. We propose to solve this problem by using a new biased pdf
\begin{equation}
q_\star(x) = (1-\xi)\,p(x) + \xi\,q(x)
\end{equation}
This expression is just a linear combination of the original pdf and the desired biased pdf, where the parameter $\xi$ sets the relative importance of the latter component. Generating a random event $x$ from the pdf $q_\star(x)$ is easy: it is a direct application of the so-called composition method or probability mixing method, a simple method that allows to generate random numbers from pdf that can be decomposed as a weighted sum 
\citep{Devroye1986, 2015A&C....12...33B}. The power of this composite biasing becomes clear if we look at the weight factor, 
\begin{equation}
w_\star(x) = \frac{p(x)}{q_\star(x)} = \frac{1}{(1-\xi) + {\xi}/{w(x)}}
\end{equation}
Even if the pdfs $p(x)$ and $q(x)$ are radically different and the weight factor $w(x)$ would become very large in some part of the domain, the composite biasing weight factor $w_\star$ is always bounded by the finite value 
\begin{equation}
w_\star(x) < \frac{1}{1-\xi}
\end{equation}
If we take $\xi=\tfrac12$, half of the events are sampled from the original pdf and the other half from the desired pdf, and the weight of a photon package will never be boosted by more than a factor two. Note that this accounts for a single event: in case a photon undergoes multiple consecutive events with composite biasing, it is possible that the cumulative weight exceeds this factor two.

In the following sections we test the advantages of composite weighting in two applications useful for (MC) radiative transfer.

\section{Emission from multiple components}
\label{MultipleComponents.sec}

\subsection{Problem description}

The first step in the life cycle of every photon package in a MCRT simulation is the generation of a random position from which it is launched into the medium. If most of the radiation originates from a point source, as is often the case for radiative transfer simulations of circumstellar discs \citep{2006ApJS..167..256R, 2013ApJS..207...30W}, reflection nebulae \citep{1977ApJS...35....1W, 1984ApJ...278..186Y} or the dusty tori around active galactic nuclei \citep{2012MNRAS.420.2756S, 2015A&A...583A.120S}, this is an easy process.

In general, the pdf that describes the position $\bfx$ from which a photon package (at a given wavelength) should be generated is the normalised monochromatic luminosity distribution,
\begin{equation}
f(\bfx) = \frac{j(\bfx)}{L}
\end{equation}
where $j(\bfx)$ is the luminosity density at $\bfx$ and $L$ is the total luminosity. In many MCRT simulations, the source that emits the radiation is built up out of a number of components. These can be a limited number of physical components (such as multiple stars, or an individual bulge and disc component in a galaxy simulation). Often, however, the source consists of a large number of components defined by the numerical setup of the simulation. A typical example is when radiative transfer is used to post-process snapshots from hydrodynamical simulations that use either grid-based or smoothed particle hydrodynamics techniques. In the former case, each cell is a component, in the latter case, each smoothed particle is a separate component. Another frequent case is the thermal emission by dust in a dust radiative transfer simulation: in such simulations, each of the grid cells that have been used to absorb the primary radiation (typically from stars or active galactic nuclei) acts as a source itself. 

Question is now how the initial position of a random photon package should be determined. If the source is composed of  $N$ different components, one can write
\begin{equation}
j(\bfx) = \sum_{m=1}^N j_m(\bfx)
\end{equation}
where $j_m(\bfx)$ is the luminosity density of the $m$'th component. The corresponding pdf from which a position should be generated is
\begin{equation}
f(\bfx) = \sum_{m=1}^N \left(\frac{L_m}{L}\right)\left(\frac{j_m(\bfx)}{L_m}\right)
\label{px-sum}
\end{equation}
with $L_m$ the total luminosity of the $m$'th component. Thanks to the composition method  \citep{Devroye1986}, the generation of a random position from the pdf (\ref{px-sum}) comes down to two steps. In a first step, a random component $m$ is chosen according to the discrete pdf
\begin{equation}
p(m) = \frac{L_m}{L}\qquad m=1,\ldots,N
\label{pm-ref}
\end{equation}
In the second step, a random position is generated from this $m$'th component by sampling a random $\bfx$ from the pdf 
\begin{equation}
f_m(\bfx) = \frac{j_m(\bfx)}{L_m}
\label{fm}
\end{equation}
In many cases the latter operation is relatively simple, in particular if the different "components" are cells with a uniform density. Generating random positions from such a pdf is almost trivial.

The problem with this approach is that the discrete pdf (\ref{pm-ref}) is often very non-uniform: the contribution to the total luminosity of the different components can easily vary many orders of magnitude (especially since we are dealing with monochromatic luminosities at one particular wavelength). The consequence is that components with a very low contribution to the total luminosity will almost never be selected, which will result in a very poor signal-to-noise in the regions where they are dominant. 

\subsection{The application of composite biasing}

We can solve this problem by biasing the pdf that dictates the choice of the component number $m$. Rather than selecting the component from the discrete pdf (\ref{pm-ref}), we can select a different discrete pdf $q(m)$, select random components from this biased pdf, and correct for the biasing by applying an additional weight to the photon package. 

A logical choice for the biased pdf would be a uniform distribution where each component has the same probability. This would guarantee that more or less the same number of photon packages is launched from each component, and thus imply a similar signal-to-noise. The biased pdf would then be
\begin{equation}
q(m)
=
\frac{1}{N}
\label{qm}
\end{equation}
with the weight
\begin{equation}
w(m)
=
\frac{p(m)}{q(m)}
=
\frac{N\,L_m}{L}
=
\frac{L_m}{\langle L \rangle}
\end{equation}
with $\langle L \rangle$ the average luminosity emitted by each component. Applying this weight to each photon package essentially comes down to assigning a luminosity to each emitted photon package that is directly proportional to the total luminosity emitted from that component. Photon packages emitted from the most luminous component obviously get the strongest boost in their luminosity. As some components typically have a much higher luminosity than others, and thus $L_m \gg \langle L \rangle$, these weight factors can become very large.

The solution is to consider composite biasing. Rather than choosing the uniform distribution (\ref{qm}) as a biased pdf, we can use 
\begin{equation}
q_\star(m) = (1-\xi)\,p(m) + \xi\,q(m) = \frac{(1-\xi)\,L_m}{L} + \frac{\xi}{N}
\label{qstarm}
\end{equation}
The weight function corresponding to this discrete biased pdf is
\begin{equation}
w_\star(m)
=
\frac{1}{(1-\xi) + \xi {\langle L \rangle}/{L_m}}
\end{equation}
The maximum weight is obviously again obtained for the photon packages emitted from the most luminous component, but the main advantage is now that the maximum weight is always limited to $1/(1-\xi)$. In this particular application of composite biasing, this weight factor can not accumulate (each photon is only emitted once).

\subsection{Implementation}

We implemented this biased emission method in the SKIRT code. SKIRT\footnote{\url{http://www.skirt.ugent.be}} \citep{2003MNRAS.343.1081B, 2011ApJS..196...22B, 2015A&C.....9...20C} is a highly modular MCRT code that comes with a suite of input models for the distribution of the stars and dust. These components can easily be altered and combined to form complex input geometries. The code also contains the framework to read in numerical input geometries defined on a hierarchical or an unstructured Voronoi grid, or as a set of SPH particles. For details, we refer to \citet{2015A&C....12...33B}. We have implemented the composite biasing as described for the primary emission (when different components are combined, as well as when the primary sources are grid or particle based) and for thermal dust re-emission. The value of the $\xi$ can be chosen by the user. 

Composite biased emission has also been implemented in DIRTY \citep{2001ApJ...551..269G, 2001ApJ...551..277M} in a very similar way.

\subsection{Simulations}

\begin{figure*}
\centering
\includegraphics[width=0.7\textwidth]{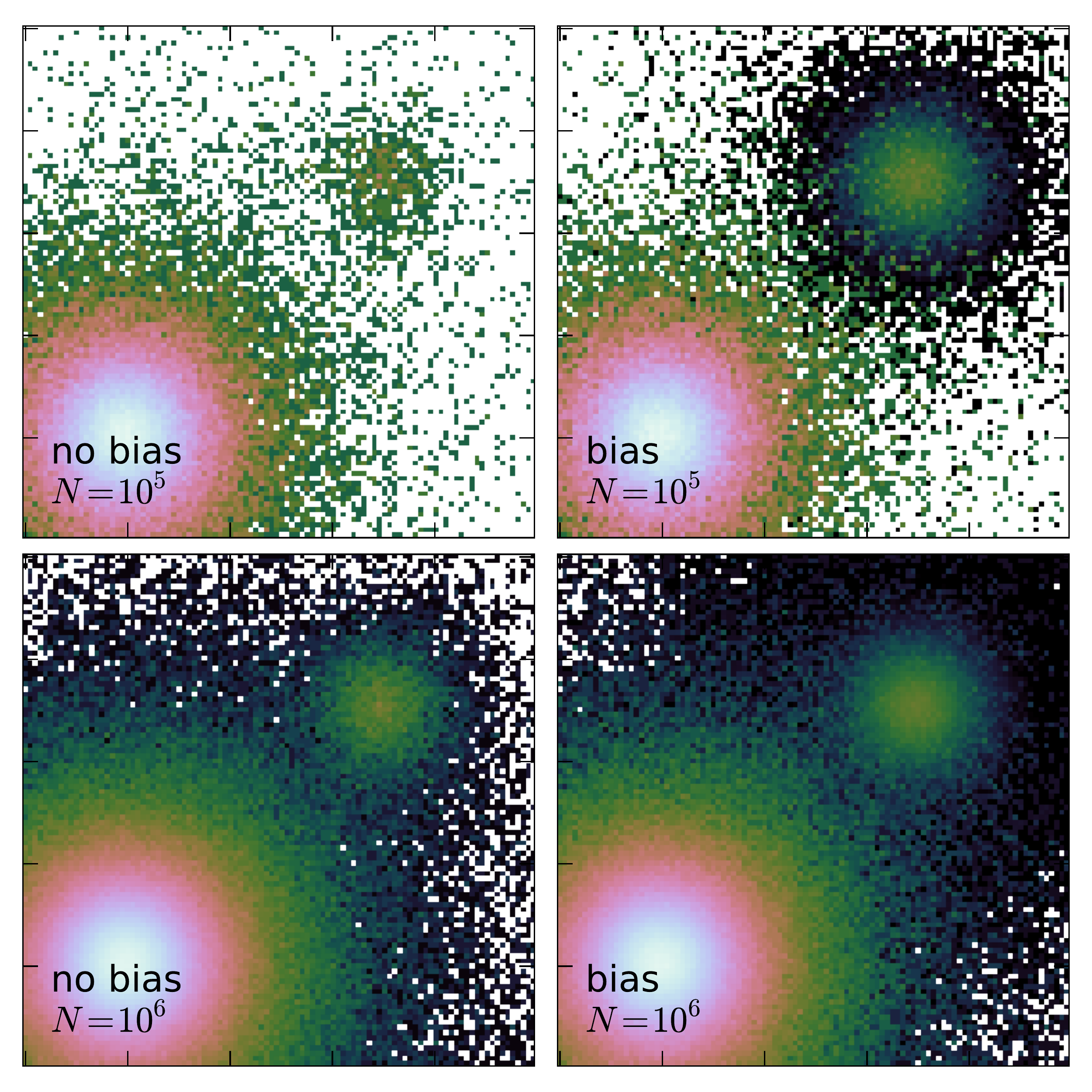}
\caption{Simulated images of a composite system composed of two Plummer spheres. The difference in intrinsic luminosity between the two components is a factor 100. The two panels on the left show the simulated image of the system without biasing (i.e.\ each component has a weight proportional to its luminosity), for $10^5$ and $10^6$ photon packages. The panels on the right show similar images for the same system, but now composite biasing is applied.}
\label{plummertwo.fig}
\end{figure*}

As a test case, we used a simple toy model consisting of two \citet{1911MNRAS..71..460P} spheres that emit radiation according to their luminosity density
\begin{equation}
j(\bfx) 
=
\sum_{m=1}^2 \frac{3L_m}{4\pi c_m^3}\left(1+\frac{|\bfx-\bfx_m|^2}{c_m^2}\right)^{-5/2}
\end{equation}
Both components have the same scale length $c_m$, but a different luminosity ($L_1/L_2 = 100$), and they are offset from each other. We ran two simulations for this system: one in which we generate initial positions of the various photon packages using the regular scheme, and one where we use the composite biasing scheme. 

The left panels of Figure~{\ref{plummertwo.fig}} show images resulting from the simulations using the regular method, with $10^5$ (top) and $10^6$ (bottom) photon packages, respectively. As this method assigns a probability to each component proportional to its luminosity, less than 1\% of all photon packages are emitted from the fainter component, and more than 99\% from the more luminous one. The result is a poor signal-to-noise for the latter component, as clearly visible in the image.

The right panels of Figure~{\ref{plummertwo.fig}} show images of the same system, but now using composite biasing with $\xi=\frac12$. For half of the photon packages, the component is selected using the same method as before (resulting in less than 1\% of these from the faint component), and the other half are selected using a heads or tails principle. The result is that 25.5\% of all photon packages are emitted by the fainter component and 74.5\% from the luminous one. The result is clearly visible in the image: the signal-to-noise of the fainter Plummer sphere is much higher compared to the previous simulation, whereas the signal-to-noise of the luminous sphere is only slightly reduced.

\begin{figure*}
\centering
\includegraphics[width=\textwidth]{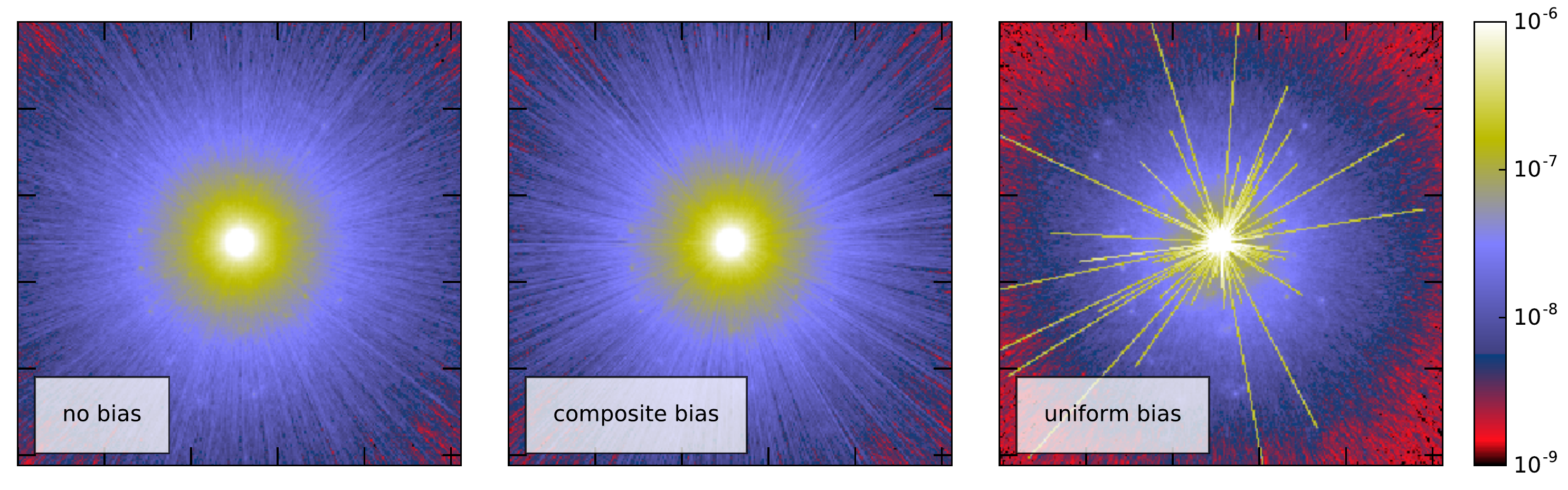}
\caption{Monochromatic mean intensity slices of a composite system composed of 1000 smoothed particles and a luminous central point source. The left panel corresponds the regular scheme (\ref{pm-ref}), the central one to the composite biasing scheme (\ref{qstarm}), and the right one to the uniform biasing scheme (\ref{qm}).}
\label{sphagn.fig}
\end{figure*}

We could have also opted for a more simple uniform pdf, and sample half of the photon packages from the luminous and half from the less luminous component. This would give both components an equal signal-to-noise. The drawback here would be that the weight difference between photon packages emitted from both components would increase. The weight of photon packages from the luminous component would be boosted by a factor 1.98, whereas those from the less luminous component would be dampened by a factor 0.020. In our composite biasing scheme, the difference between these two is significantly lower: the weight factor for the luminous component is only 1.33, whereas the one for the less luminous one is 0.039.

It is easy to come up with more extreme examples. In Figure~{\ref{sphagn.fig}} we consider a system composed of 1000 emitting smoothed particles, with positions randomly generated from a uniform density sphere, and an additional central point source. Each of the smoothed particles has the same luminosity, and the point source is 1000 times more luminous than a single smoothed particle. This could be an idealised simulation of an active galactic nucleus (AGN) embedded in a smooth stellar system. Figure~{\ref{sphagn.fig}} shows the mean intensity of the radiation field in a 2D slice through the simulation volume, calculated using the method of \citet{1999A&A...344..282L}. The three panels correspond to different schemes: without biasing (left), with composite biasing (centre) and with uniform biasing (right). In the simulation with the regular scheme, half of the photon packages are emitted from the nuclear point source, while the other half are distributed over all the smoothed particles, which results in a relatively poor signal-to-noise in the observed image. If we adopt uniform biasing and assign the same probability to each component, the vast majority of the photon packages are emitted by the smooth particles and the signal-to-noise in the image is improved. However, the few photon packages emitted by the AGN component are boosted by a factor of more than 500, and this has a huge effect on the mean intensity. The photon packages emitted from the AGN have very large weights and they leave a bright trace through the system, and boost the value of the mean radiation field in the cells they pass through to unrealistically high values, whereas the level of the radiation field strength is underestimated in those cells that are not crossed by an AGN photon package. The composite biasing scheme resides between these two extremes: the signal-to-noise of the host galaxy is better than in the regular scheme, whereas the weight factor for photon packages emitted by the AGN is strongly reduced to slightly below 2, which results in a mean intensity distribution that is not affected by extreme boosting.

The two example source geometries, shown in Figs.~{\ref{plummertwo.fig}} and {\ref{sphagn.fig}}, illustrate an important point concerning composite biasing. In the first case, choosing the regular method was a poor idea, since it produced a poor signal-to-noise for the second component. We described the benefit of a biasing scheme with composite biasing, but in the end, also a simple uniform pdf would not have been a bad option. For the second case, biasing with a uniform pdf would have been a poor choice because of the huge weight factor, whereas the regular scheme would be a reasonable alternative to our composite biasing scheme. In both cases, we could hence obtain a good simulation without our composite biasing scheme, but we would have needed to pick a different algorithm for each situation. This is, in our opinion, a nice advantage of the composite biasing scheme (at least for its application to emitting photon packages from multiple components): it is one single recipe that can be applied to these two radically different cases. There is hence no need to study the characteristics of the radiative transfer problem in detail before choosing a proper algorithm.

\section{Path length stretching}
\label{PathlengthStretching.sec}

\subsection{Problem description}

We now turn our attention to a different problem where biasing, and composite biasing in particular, can contribute, namely the problem of penetration of radiation through an optically thick medium. This is a notoriously hard nut to crack for MCRT simulations 
\citep[e.g.,][]{2009A&A...497..155M, 2009A&A...498..967P}.

The standard radiative transfer theory \citep[e.g.,][]{1960ratr.book.....C} dictates that the pdf corresponding to the path length covered between two interaction events (i.e., emission, absorption or scattering events) in optical depth coordinates is an exponential distribution. In an MCRT code this means that, after the position and propagation direction of a photon package have been determined, a random optical depth $\tau$ is generated from an exponential distribution,
\begin{equation}
p(\tau) = \txe^{-\tau}
\label{ptau}
\end{equation}
This optical depth is subsequently converted to a physical path length, which can be combined with the initial position and propagation direction to yield the next interaction site. 

It is immediately clear why large optical depths pose a challenge for MCRT simulations. An exponentially declining distribution is strongly peaked towards small values, and the probability that a large optical depth is randomly generated from such a distribution is hence extremely small. For example, the probability that a value $\tau>20$ is chosen is only one out of 485 million, and for $\tau=50$ this reaches one out of $5.2\times10^{21}$. If we want to determine the radiation field at a position behind an optically thick barrier, we hence have to generate huge numbers of photon packages to ensure that some of them have a random optical depth large enough to pass the barrier. This would result in an excessive computation time. 

\subsection{The application of composite biasing}

\begin{figure*}
\centering
\includegraphics[width=\textwidth]{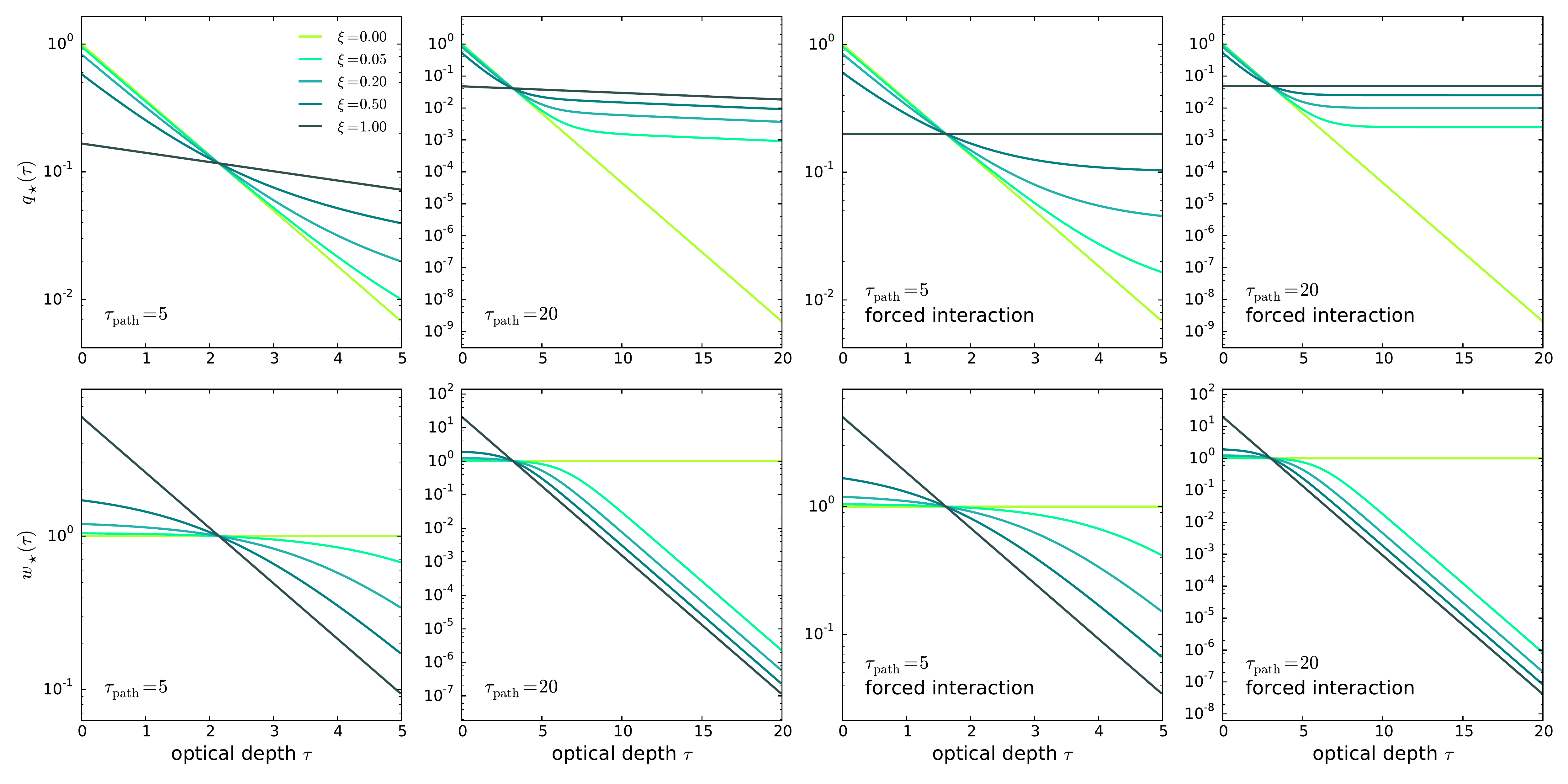}
\caption{The composite biased pdf (top row) and the corresponding weight factor (bottom row) for path length stretching. The first and second column correspond to the path length stretching for two different optical depths along the path ($\tau_{\text{path}} = 5$ and 20 respectively). The third and fourth columns correspond to the same cases, but now path length stretching is combined with forced interaction. Within each panel, the different lines correspond to different values of the mixing parameter $\xi$, ranging from $\xi=0$ (no biasing) to $\xi=1$ (pure biasing, no composite biasing).}
\label{stretching.fig}
\end{figure*}

One way to tackle this problem is by biasing the pdf (\ref{ptau}), such that larger values of the optical depth are artificially promoted compared to lower values. This idea has mainly been used in the nuclear engineering community, where it is known under the names path length stretching or exponential biasing \citep{Levitt1968, Spanier1970, Dwivedi1982}. In these studies, the exponential pdf (\ref{ptau}) is replaced by 
\begin{equation}
q(\tau) = \alpha\,\txe^{-\alpha\tau}
\label{qtau}
\end{equation}
This means that, effectively, the cross section or opacity of the medium is modified by a factor $\alpha$. The stretching parameter $\alpha$ can be given any value between 0 and 1 to provide a more uniform sampling, and hence to promote the generation of higher optical depth values. The weight factor corresponding to the biased pdf (\ref{qtau}) is
\begin{equation}
w(\tau)
= 
\frac{{\text{e}}^{-(1-\alpha)\,\tau}}{\alpha}
\label{wtau}
\end{equation}
which, logically, has its maximum for $\tau=0$,
\begin{equation}
w_{\text{max}} = \frac1\alpha
\label{wmax1alpha}
\end{equation}
A crucial issue is to determine an optimal value for $\alpha$. The simplest option would be to consider a single uniform value, but it seems more reasonable to fine-tune the value of $\alpha$ depending on $\tau_{\text{path}}$, the total optical depth between the current position and the boundary of the model space.\footnote{In some codes, the value of $\tau_{\text{path}}$ is calculated after every scattering/emission event before the next interaction site is randomly generated. This is particularly the case for those codes where continuous absorption is implemented along the entire ray \citep{2003A&A...399..703N, 2011ApJS..196...22B}. For such codes, the implementation of a scheme where $\alpha$ depends on $\tau_{\text{path}}$ does not imply any overhead. Other MCRT codes do not automatically calculate $\tau_{\text{path}}$ along every path and calculate the next interaction point on-the-fly by integrating along the path until the covered optical depth reaches the randomly determined value. In this case, the implementation of such a scheme would imply some additional calculations.} 
Indeed, for paths with a modest optical depth, $\tau_{\text{path}} \lesssim 1$, no biasing or soft biasing with $\alpha\lesssim1$ is sufficient. For paths with a large optical depth, $\tau_{\text{path}} \gg 1$, a much stronger biasing with $\alpha\ll1$ is required in order to promote the probability of reaching the outer regions of the path. A simple choice that would guarantee that the outer part of the path is always promoted in roughly the same way is 
\begin{equation}
\alpha(\tau_{\text{path}}) = \frac{1}{1+\tau_{\text{path}}}
\label{alphataupath}
\end{equation}
The combination of equations (\ref{wmax1alpha}) and (\ref{alphataupath}) results in weight factors that can grow relatively large, in particular in simulations with a large optical depth. In other words, in the regime of large optical depths, where we need to choose $\alpha$ small enough to promote sufficient penetration to the outer regions along the path, we run the risk of creating photon packages with large weight factors.

Composite biasing (or composite path length stretching) presents a solution to this problem: instead of the biased pdf (\ref{qtau}) we adopt
\begin{equation}
q_\star(\tau)
=
(1-\xi)\,{\text{e}}^{-\tau} + \xi\,\alpha\,{\text{e}}^{-\alpha\tau}
\label{qstartau}
\end{equation}
with corresponding weight factor
\begin{equation}
w_\star(\tau)
= 
\frac{1}{(1 - \xi) + \xi\,\alpha\,{\text{e}}^{(1-\alpha)\,\tau}}
\label{wstartau}
\end{equation}
This weight factor never grows larger than 
\begin{equation}
w_{\star,\text{max}} = \frac{1}{1 - (1-\alpha)\,\xi} < \frac{1}{1-\xi}
\end{equation}

The panels on the two leftmost columns in Figure~{\ref{stretching.fig}} show the biased pdf $q_\star(\tau)$ and the corresponding weight function $w_\star(\tau)$ for two different values of $\tau_{\text{path}}$, where we have used the relation (\ref{alphataupath}). In each panel, the different curves correspond to different values of the parameter $\xi$, with $\xi=0$ corresponding to no biasing, and $\xi=1$ to complete biasing (i.e., no composite biasing). In each case, the weight function decreases monotonically with increasing optical depth. The effect of composite biasing is clearly visible: complete biasing reshuffles the original pdf drastically, with the implication that the weight factor can assume large values for small optical depths (especially for larger values of $\tau_{\text{path}}$). Composite biasing creates a mixture of the original and desired pdf; in particular, it has the same slope as the original pdf for small $\tau$ and the same slope as the desired pdf for large $\tau$. The result is that weight function remains under control at small optical depths.

Interestingly, all curves in each panel pass through a single point, located at 
\begin{equation}
\tau_{\text{crit}} = \frac{-\ln\alpha}{1-\alpha}
\end{equation}
This critical optical depth corresponds to $w_\star=1$, and it hence divides the range of optical depths in a part where the weight factor is boosted  ($\tau<\tau_{\text{crit}}$) and a part where it is dampened ($\tau>\tau_{\text{crit}}$).

\subsection{Combination with forced interaction}

A biased pdf of the form (\ref{qtau}), i.e.\ a shallower exponential pdf, is obviously not the only possibility. A very interesting option to promote different sections along the path more equally is the combination of composite path length stretching and forced interaction. Forced interaction or forced scattering \citep{Cashwell1959, 1977ApJS...35....1W} is an MCRT acceleration technique originally developed to deal with radiative transfer in the low optical depth regime. It prevents photon packages to leave the system without an interaction. It is by itself an application of the biasing technique, where the exponential distribution (\ref{ptau}) is cut off at $\tau_{\text{path}}$. This forced interaction is corrected for by changing the weight of the photon package by an appropriate weight factor
\begin{equation}
w_{\text{fi}}
=
1-{\text{e}}^{-\tau_{\text{path}}}
\label{wfi}
\end{equation}
Among the existing MCRT codes, different strategies are adopted concerning forced interactions: some codes only force the first interaction after the emission of each photon package, and transition to unforced interaction for the remainder of the life cycle of the photon package (in this case, the name forced first scattering is adopted). Other codes apply forced interaction throughout the simulation, an approach sometimes called eternal forced interaction.

Now assume that we one wants to apply forced interaction. The first step is then to apply the weight factor~(\ref{wfi}) to the photon package, and the second step is to generate a random $\tau$ from the appropriately normalised pdf 
\begin{equation}
p(\tau) 
= 
\frac{{\text{e}}^{-\tau}}{1-{\text{e}}^{-\tau_{\text{path}}}}
\qquad 
0 \leqslant \tau < \tau_{\text{path}}
\label{ptau-fi}
\end{equation} 
If the total optical depth to the model boundary is small ($\tau_{\text{path}}<1$), this pdf is close to a uniform distribution, which implies that all values of $\tau$ have more or less equal probability to be generated. For large values of $\tau_{\text{path}}$, the distribution is strongly peaked and it is very unlikely that a large $\tau$ will be chosen. 

We can of course also apply path stretching in this case, i.e.\ we can bias the pdf (\ref{ptau-fi}) in a similar way as we did to the pdf (\ref{ptau}). Since this pdf is defined on a domain with a finite range, we have a different set of biased pdfs that we can consider. An obvious candidate would be a uniform distribution,
\begin{equation}
q(\tau) = \frac{1}{\tau_{\text{path}}}
\qquad 
0 \leqslant \tau < \tau_{\text{path}}
\label{qtau-fi}
\end{equation}
When this option is chosen, there is an equal probability (in optical depth space) for the scattering to take place anywhere along the path to model boundary. The problem with this choice is the weight factor
\begin{equation}
w(\tau) 
= 
\frac{\tau_{\text{path}}\,{\text{e}}^{-\tau}}
{1-{\text{e}}^{-\tau_{\text{path}}}}
\end{equation}
which can grow very large in the regime of high optical depths,
\begin{equation}
w_{\text{max}} 
= 
\frac{\tau_{\text{path}}}{1-{\text{e}}^{-\tau_{\text{path}}}}
\end{equation}
This problem is avoided by using composite path length stretching,
\begin{equation}
q_\star(\tau) 
= 
\frac{(1-\xi)\,{\text{e}}^{-\tau}}{1-{\text{e}}^{-\tau_{\text{path}}}} 
+ \frac{\xi}{\tau_{\text{path}}}
\qquad 
0 \leqslant \tau < \tau_{\text{path}}
\label{qstartau-fi}
\end{equation} 
with the associated weight function
\begin{equation}
w_\star(\tau) 
= 
\frac{1}
{(1-\xi)
+\xi\left(\frac{1-{\text{e}}^{-\tau_{\text{path}}}}{\tau_{\text{path}}}\right){\text{e}}^{\tau}}
\label{wstartau-fi}
\end{equation}
The maximum weight is now 
\begin{equation}
w_{\star,\text{max}} 
= 
\frac{1}
{(1-\xi)
+\xi\left(\frac{1-{\text{e}}^{-\tau_{\text{path}}}}{\tau_{\text{path}}}\right)}
<
\frac{1}{1-\xi}
\end{equation}

\begin{figure}
\centering
\includegraphics[width=0.3\textwidth]{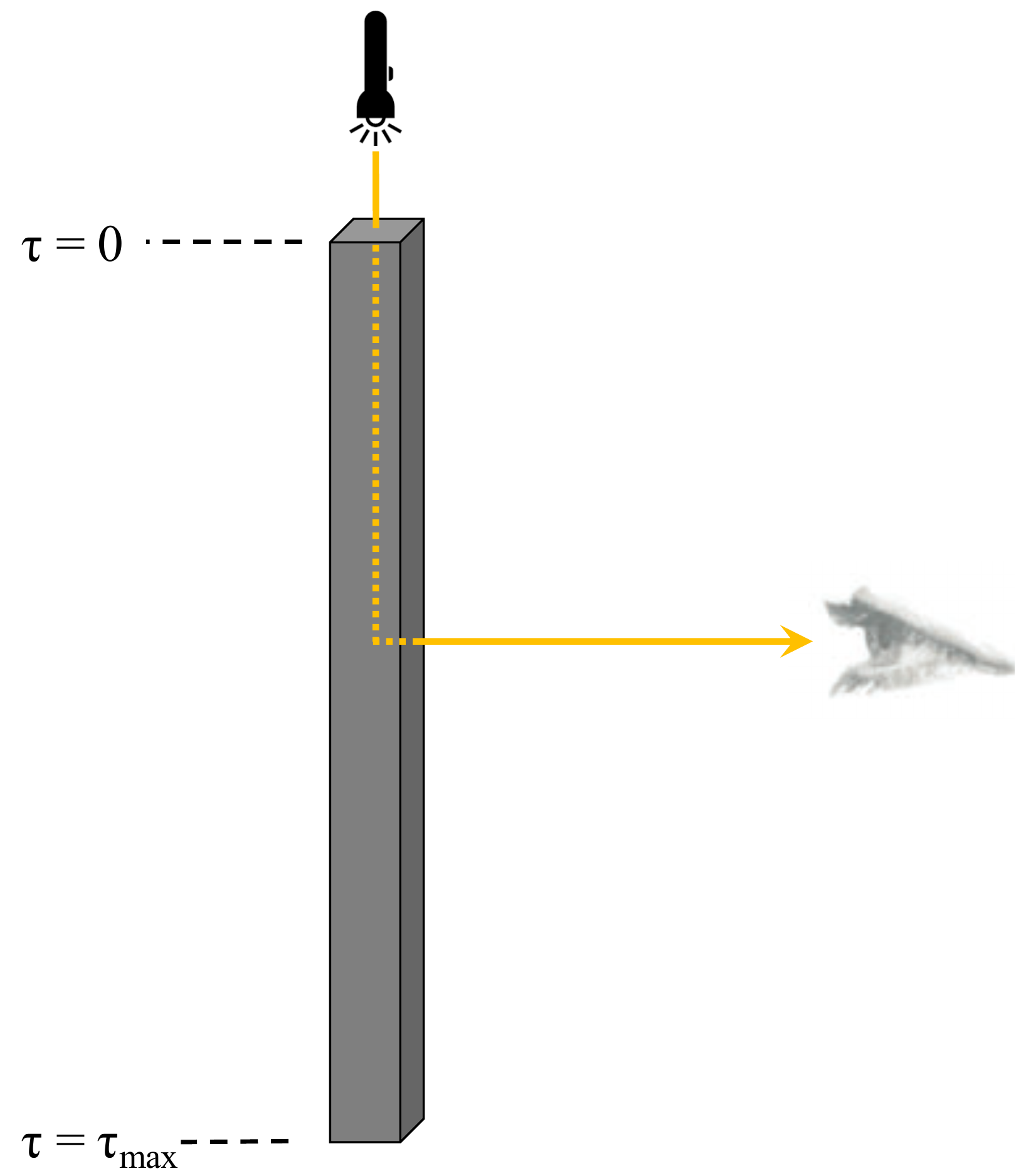}
\caption{The geometry of the uniform density pillar model described in Section~{\ref{slabbysim.sec}}. }
\label{slabbygeometry.fig}
\end{figure}

\begin{figure*}
\centering
\includegraphics[width=\textwidth]{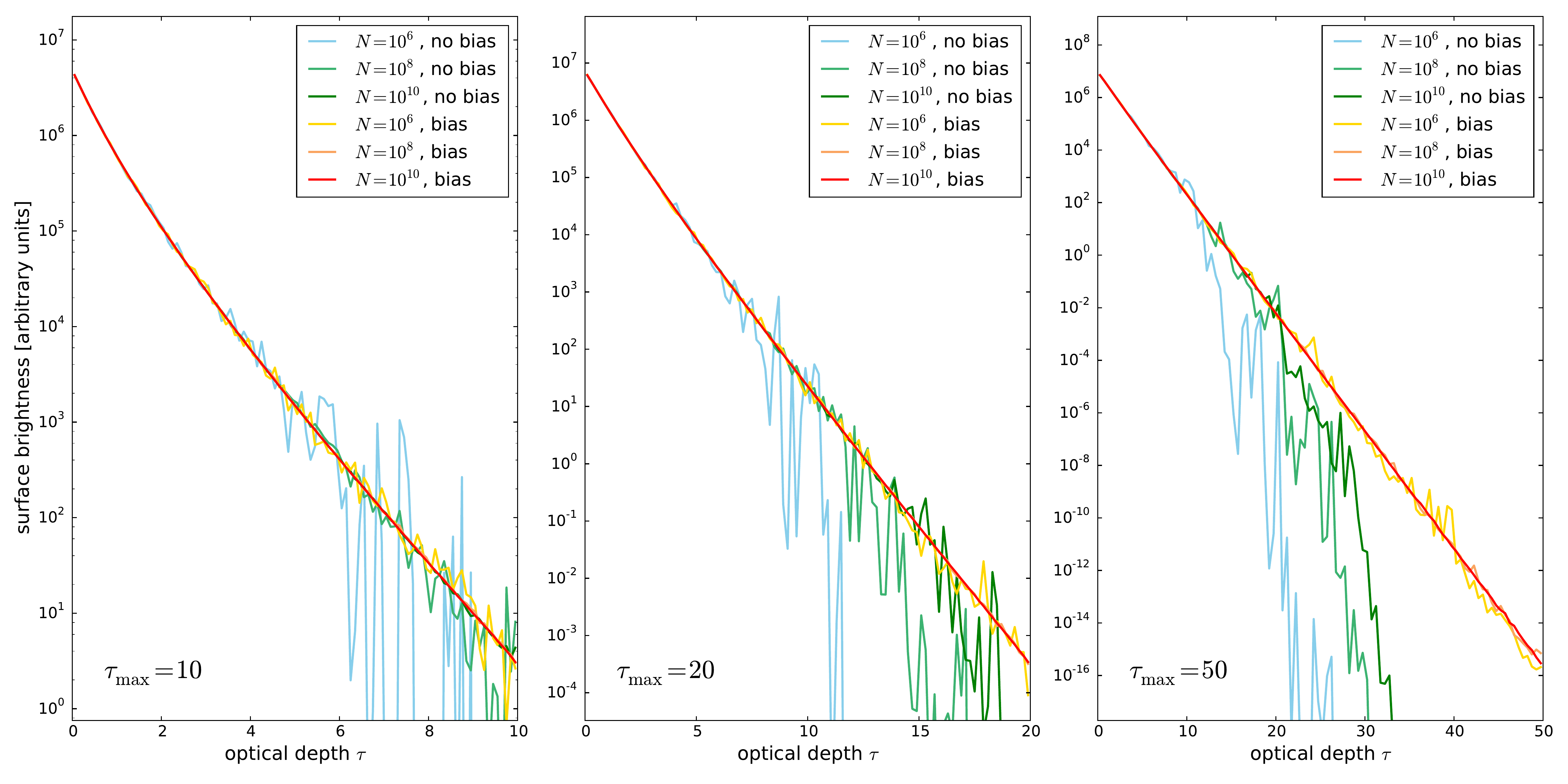}
\caption{Vertical surface brightness profiles of the simple pillar model described in Section~{\ref{slabbysim.sec}}. The three panels correspond to three different values of the total optical depth through the pillar. The horizontal axis indicates the depth through the pillar, measured from the top where the radiation enters. The different lines in each panel correspond to models with a different number of photon packages (ranging from $10^6$ to $10^{10}$) and with and without the application of composite path length stretching.}
\label{slabby.fig}
\end{figure*}

The panels on the two rightmost columns in Figure~{\ref{stretching.fig}} are similar to the leftmost panels, but now correspond to the case where composite biasing is combined with forced interaction. The same general conclusions can be drawn. Also in this case, the different curves in each panel pass through a single point, now located at
\begin{equation}
\tau_{\text{crit}} = -\ln\left(\frac{1-\txe^{-\tau_{\text{path}}}}{\tau_{\text{path}}}\right)
\end{equation}
Again, this critical optical depth corresponds to $w_\star=1$, and hence divides the range of optical depths in a boosting and a damping part.

\subsection{Implementation}

We implemented composite path length stretching in the SKIRT code. As the code adopts eternal forced interaction, we implemented the biased pdf (\ref{qstartau-fi}), with a value of  $\xi$ to be chosen by the user. The implementation is almost trivial: the only changes required are the replacement of optical depth generation and the multiplication of the weight of the photon packages by an extra weight factor. Actually, no more than 10 lines of code were altered on a total of more than 50,000 lines in the current version of SKIRT.

While all test performed in this paper were run with SKIRT, composite path length stretching was also implemented in DIRTY, TRADING \citep{2008A&A...490..461B}, and CRT \citep{2012A&A...544A..52L}.

\subsection{Simulations and results}
\label{slabbysim.sec}

In order to test our new recipe, we designed a simple test case (Figure~{\ref{slabbygeometry.fig}}). The model consists of an cuboidal uniform density pillar of dimensions $1\times1\times10$, illuminated by a single anisotropic point source located on top of it, emitting in the negative $z$ direction. The observer is located far from the pillar and observes it from the side. This setup is optimised to study the penetration of radiation to various optical depths. This model is scale-free, and completely defined by the vertical optical depth through the pillar, $\tau_{\text{max}}$ and the optical properties of the medium. We adopt a scattering albedo of 0.5 and isotropic scattering, and consider different values of $\tau_{\text{max}}$.

The left panel of Figure~{\ref{slabby.fig}} compares the vertical surface brightness profile of the model with $\tau_{\text{max}} = 10$. The blue, green and dark green lines correspond to models without path length stretching, for different photon package numbers in the simulation. In all cases, the surface brightness in the top section of the pillar ($\tau\lesssim1$) is well determined, and the agreement between the different runs is nearly perfect. This is not surprising as this is where the radiation enters the system. The situation gets gradually worse the deeper we follow the radiation into the pillar. For the simulation with only $10^6$ photon packages, the surface brightness profile becomes noisy around $\tau\sim4$ and completely breaks down from $\tau\sim6$. There are just no photon packages that penetrate this deeply.

\begin{figure*}
\centering
\includegraphics[width=0.9\textwidth]{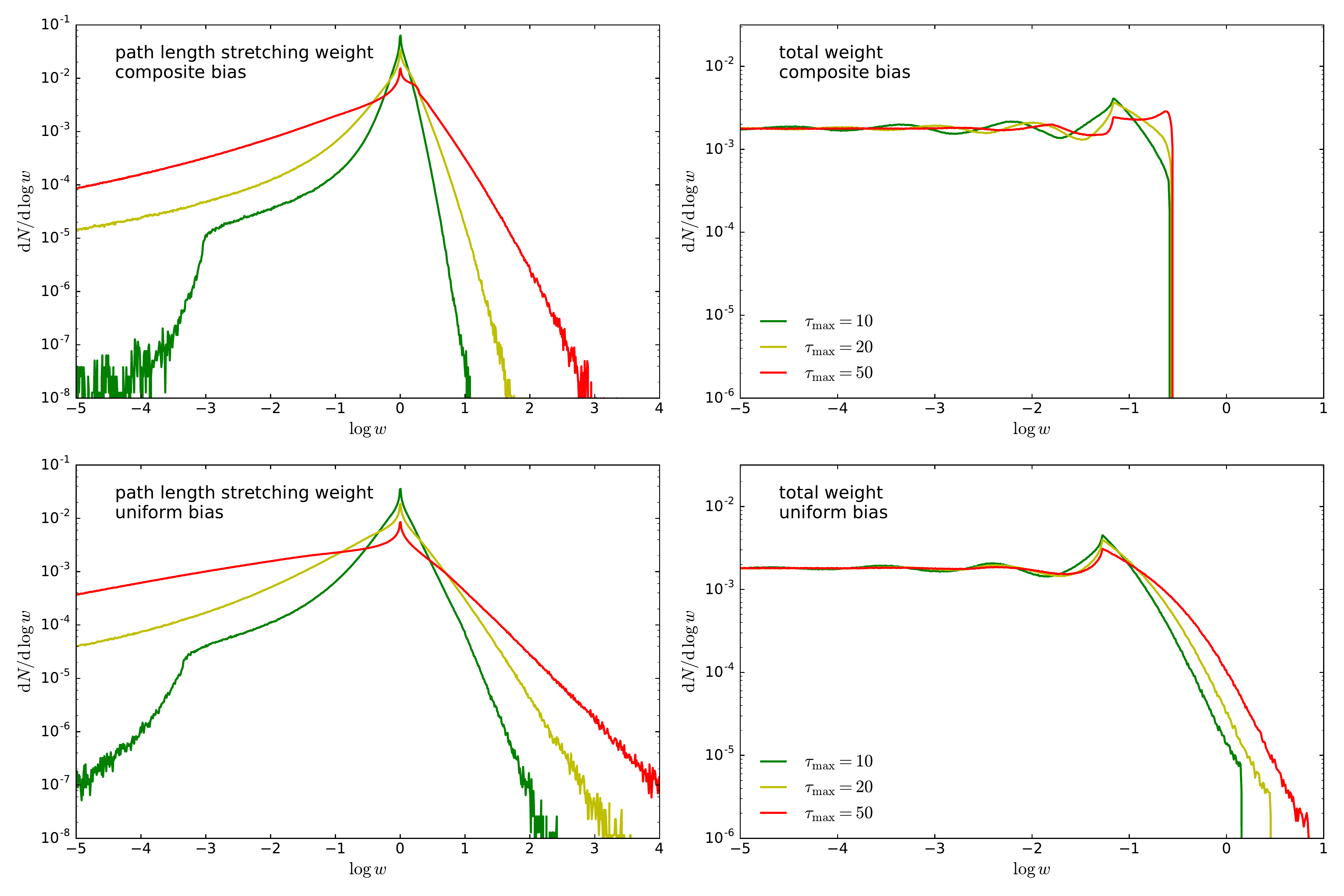}
\caption{The distribution of the cumulative weight factor of all photon packages in the pillar simulations described in Section~{\ref{slabbysim.sec}}. The panels in the left column correspond to the cumulative weight factor contribution only due to multiple applications of path length stretching. The panels on the right correspond to the total weight factor, i.e.\ also including the contribution from other events that alter the weight of a photon package. The top row shows the results of simulations with the composite biasing scheme (\ref{qstartau-fi}), the bottom panels correspond to the uniform biasing scheme (\ref{qtau-fi}).}
\label{whisto.fig}
\end{figure*}

\begin{figure*}
\centering
\includegraphics[width=0.65\textwidth]{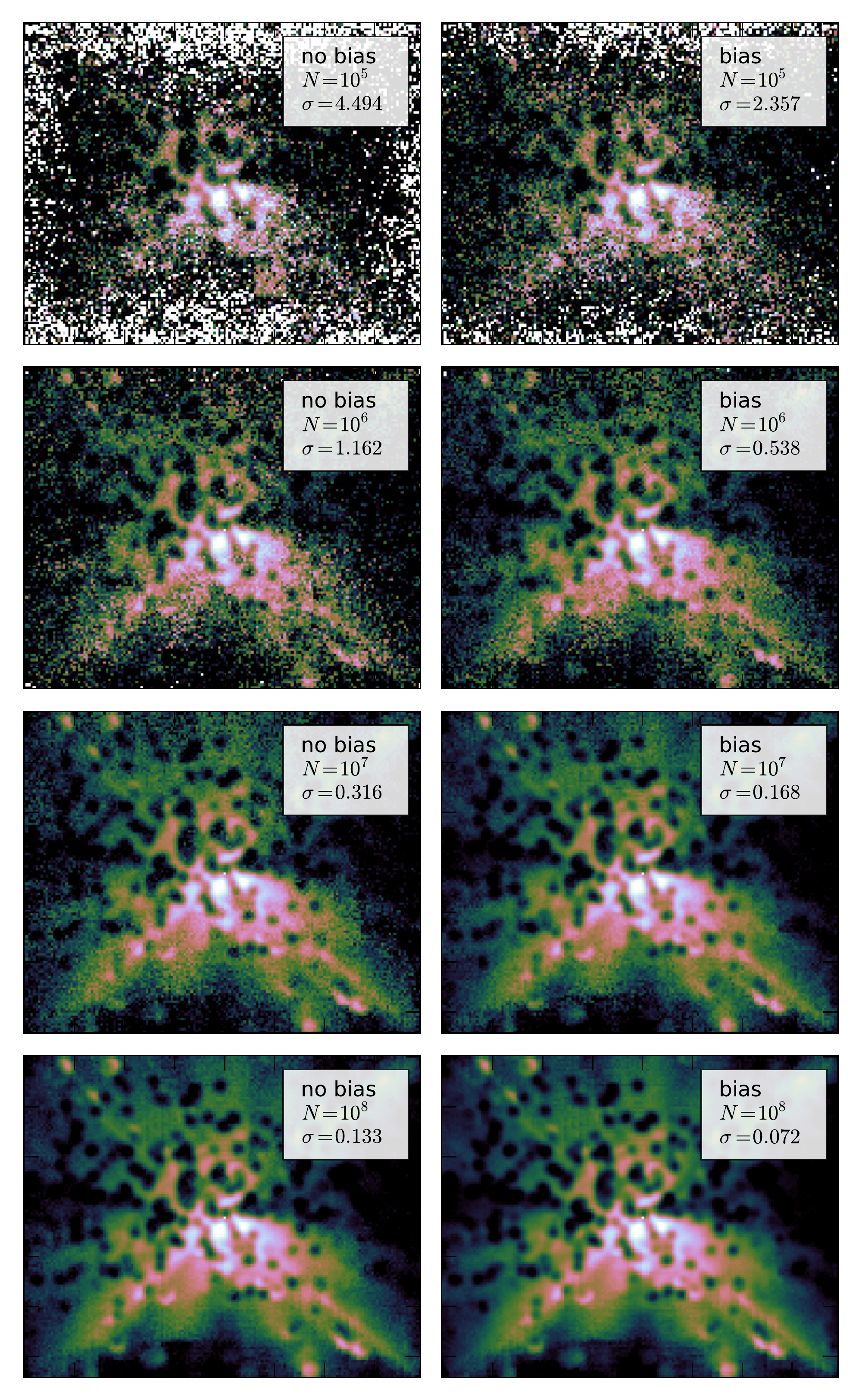}
\caption{Simulated images of the clumpy torus model described in Section~{\ref{slabbysim.sec}}. The panels in the left column correspond to simulations without biasing, those in the right column include composite path length stretching. The number of photon packages used in the simulation increases from the top row to the bottom row. The estimated average noise level in each image is indicated in each panel.}
\label{torus.fig}
\end{figure*}

When the number of photon packages in the simulation increases, the signal-to-noise of the surface brightness profile obviously increases, but also the penetration depth into the pillar increases. Indeed, with more photon packages, the probability that at least one of them has a value of $\tau$ deep in the tail of the exponential distribution becomes larger. With $10^{10}$ photon packages, photon packages can penetrate the entire pillar and the solution converges. Adding more photon packages further increases the signal-to-noise, but does not systematically alter the solution anymore.

The yellow, orange and red lines correspond to simulations with our composite path length stretching algorithm (with $\xi=\tfrac12$). The three simulations were run with the same number of photon packages as for the simulations without path length stretching. Even with only $10^6$ photon packages, the surface brightness profile is already well reproduced, and the simulation with $10^8$ photon packages produces a very smooth surface brightness profile that covers the entire pillar down to $\tau=10$, and that agrees very well with the non-biased result using $10^{10}$ photon packages. This demonstrates not only the accuracy, but also the efficiency of our approach.

The central and right panels of Figure~{\ref{slabby.fig}} show similar simulations, but now with optical depths of 20 and 50, respectively. Similarly to the $\tau_{\text{max}}=10$ case, the surface brightness profile is well determined in the top layer for all simulations, as expected. However, it gets progressively noisier when we move deeper into the pillar for the non-biased simulations, until there is no signal anymore. When the number of photon packages in the simulation increases, the penetration depth increases, as the tail of the exponential distribution gets better sampled. However, it is clear that a huge number of photon packages is necessary. For the $\tau_{\text{max}}=20$ simulation, even a simulation with $10^{10}$ photon packages does not yet fully penetrate the entire pillar, and for the $\tau_{\text{max}}=50$ simulation, the situation is hopeless for any reasonable number of photon packages. On the other hand, in our simulations with composite path length stretching, the photon packages can easily penetrate through the entire pillar, resulting in a high signal-to-noise surface brightness profile even for a modest number of photon packages. 

A particular aspect of the application of composite biasing to path length stretching is that photon packages are scattered multiple times, and hence that the weight factor (\ref{wstartau-fi}) is applied multiple times. As a result, the effective or cumulative weight factor can grow to values exceeding two. This is illustrated in the top left panel of Figure~{\ref{whisto.fig}}, where we plot the distribution of the cumulative weight factor due to path length stretching for all photon packages in the pillar simulations. These distributions show a strong peak at $w=1$, as expected. Some photon packages, however, can reach large weight factors, even up to $10^3$. This seems alarming, as this is actually what we wanted to avoid using composite biasing in the first place. However, the situation is not as bad as it seems at first sight. Indeed, the weight factor contribution due to path length stretching can only grow to very large values after many scattering events. Between two such events, the weight factor is automatically dampened by other processes, in particular by forced interaction (equation \ref{wfi}) and the scattering-absorption split \citep{2013ARA&A..51...63S}. The net result is that the total cumulative weight factor of photon packages always remains limited, as shown in the top right panel of Figure~{\ref{whisto.fig}}.

The bottom panels of Figure~{\ref{whisto.fig}} show similar distributions, but now for the uniform biasing scheme~(\ref{qtau-fi}). The bottom left panel shows that the weight factor distribution due to path length stretching is significantly broader, since each individual scattering event can result in a large weight factor. For the $\tau_{\text{max}} = 50$ case, cumulative weights larger than $10^4$ can be obtained. The bottom right panel shows that, contrary to the composite biasing scheme, also the total weights can grow large. This again shows the advantage of the composite biasing approach.

We have also considered different simulations with a more complex geometry. Figure~{\ref{torus.fig}} shows the results of a set of simulations of a clumpy torus model, similar to the AGN torus models used by \citet{2012MNRAS.420.2756S} and \citet{2013A&A...554A..10S}. Half of the obscuring material in the torus is distributed in a smooth component in which the density decreases as $r^{-1}$ in the radial direction, and the remaining half is locked up in $10^3$ optically thick clumps. The effective optical depth of the model (i.e., the optical depth if all the material were smoothly distributed) in the equatorial plane is set to 20, but due to the clumpiness and the density gradients, there are strong differences from path to path. 

The panels in the left column of Figure~{\ref{torus.fig}} show the result of SKIRT simulations without application of the biasing technique. Each panel shows the same view of the torus from an inclination angle of 80 degrees, with the number of photon packages in the simulation increasing from $10^5$ to $10^8$. The right column panels show the corresponding simulations including composite path length stretching with the same number of photon packages. In both columns, the signal-to-noise in the images logically increases when the number of photon packages increases. Comparing the left and right panels for the same number of photon packages, one can see that the simulations without biasing have a lower signal-to-noise compared to the images with biasing, especially in the outer regions. We have quantified the noise in each image in the following way. For each pixel, we calculated the relative difference between the surface brightness in the image, and the surface brightness in a very high signal-to-noise reference image (we used the mean of two simulations with $N=10^9$, one with and one without path length stretching). We then calculated the standard deviation $\sigma$ of this distribution of relative differences. The results are indicated in each panel. For each fixed value of $N$, the noise level is systematically roughly a factor two lower for the image with biasing compared to the one without. For $N=10^5$, the simulation without biasing has a noise value of 4.494, compared to 2.357 for the one with composite biasing. For $N=10^8$ the noise levels decrease to 0.133 and 0.072 respectively.

\section{Discussion and conclusions}
\label{Conclusions.sec}

Biasing is a powerful technique in MC calculations in general, and in MCRT simulations in particular. One of the beautiful aspects of biasing is the freedom one has in the choice of the biased pdf: in principle one can use any possible pdf, as long as one corrects for it by applying the correct weight factors. One of the drawbacks of the use of biasing is the potential introduction of large weight factors, especially if the chosen biased pdf differs radically from the original pdf. We have discussed a general strategy to suppress the appearance of large weight factors: we propose a linear combination of the original and the biased pdf. This combination ensures that the maximum boost factor for a photon package never exceeds a preset limit.

We have discussed in this paper two applications of biasing in MCRT that are not commonly used in current state-of-the-art MCRT codes. 

Our first application deals with the generation of random positions from an emitting source that consists of multiple components. As described in Section~{\ref{MultipleComponents.sec}}, this is a very common case, and the number of components can range from only a few to several millions. In some cases, it makes sense to give equal weight to each of these components, irrespective of their contribution to the total luminosity, whereas in other cases, it is more logical to give each component a weight proportional to its contribution to the total luminosity. Composite biasing, where these two extremes are averaged out, can be used as way to satisfy both of these demands. It can hence be used without deep a priori knowledge of the system to be modelled. This is extremely valuable when one needs to run many models in an automated way and one does not have the opportunity to study the properties of each configuration in detail. It is particularly useful when a large library of radiative transfer models is created \citep[e.g.,][]{2006ApJS..167..256R, 2012MNRAS.420.2756S, 2015A&A...583A.120S}, or when radiative transfer models are fit to observational data \citep[e.g.,][]{2012ApJ...746...70S, 2014MNRAS.441..869D, 2015A&A...579A.103V}. 

Our second application is an extension of the so-called path length stretching, a technique that has been explored in the nuclear engineering community. It is designed to bias the distribution of step lengths between two interactions in such a way that steps corresponding to larger optical depths are more easily chosen. A simple test problem demonstrates that our composite biasing method helps to penetrate high optical depth barriers, which always pose a strong challenge for MCRT codes. The implementation of the algorithm is trivial, and it does not interfere with any of the other optimisation techniques built in into most MCRT codes, such as biased emission, continuous absorption or photon-package peel-off \citep{2013ARA&A..51...63S}. For our most simple test case, it corresponds to a gain in efficiency of at least three orders of magnitude (in the sense that simulations with at least 1000 times more photon packages are necessary to obtain a surface brightness profile with a comparable signal to noise). For higher optical depths, the gain is many orders of magnitude more. 

Both applications that we have explored demonstrate the benefit of biasing, and composite biasing in particular. They are definitely not the last word on biasing that can (and needs to) be said, however. 

First of all, all MCRT algorithms involve the sampling of many different probability density functions. We have just dealt with two of these, but there are probably others that one could consider for biasing in order to gain efficiency. In this context, it is striking that some techniques, such as the path length stretching algorithm, have become relatively mainstream in some MC communities and remain completely unknown in others. More cross-fertilisation across domain borders is hence strongly recommended.

Second, we have proposed composite biasing prescriptions for two selected cases, but we do not make the claim that these prescriptions are optimal. This is particularly relevant for the path length stretching case. In the setup where path length stretching is combined with forced interaction, we have adopted a biased optical depth distribution that is the weighted sum of a uniform and an exponential distribution. This biased distribution has the advantage that it can nicely cover the entire optical depth range, while still avoiding the creation of photon packages with extreme weights. Still, this choice might not always be optimal. Consider, for example, an extreme case consisting of a compact, very optically thick region, embedded within an extended low-density medium. In the standard approach with an exponentially declining pdf, the interaction point will almost certainly be located in the dense region. If we use our biased pdf, the possible interaction locations will be spread out much more in optical depth space, but they will still cluster within the same small physical region. The result is that it will remain hard for a photon package to escape the high optical depth region. Our biased approach will hence not solve the problem in a significant way (but, on the other hand, it will also not perform significantly worse compared to the unbiased method). For this case, it would make more sense to consider a different form for the biased pdf. Rather than a uniform pdf in optical depth space, one could consider a uniform pdf in physical path length space, or a discrete pdf that gives equal weight to each cell crossed along the path. Obviously, one could consider any mixture of these -- the possible variations are endless.

Finally, even though our example simulations suggest that our composite biasing schemes for the emission from multiple components and for path length stretching are giving accurate results and making simulations more efficient, one should recall that biasing remains a give-and-take. One simply cannot gain on every front: if one promotes certain parts of the domain of a pdf, one automatically demotes other parts. The usual goal of biasing is to redistribute noise in a clever way, and reduce the signal-to-noise in parts where it is high anyway, and invest that in parts where the signal-to-noise is unacceptably low. If, however, one is only interested in the most luminous parts of a system and one cares less about the fainter components, biasing could be less beneficial or even counter-productive. 

Whether or not biasing is beneficial is not always obvious to predict. For example, we have discussed two examples in which path length stretching is an effective way to penetrate optically thick regions, but it is not necessarily a solution to all radiation transfer problems with high optical depths. Especially cases where most of the signal is caused by photon packages that have been scattered multiple times are challenging even with path length biasing. If path stretching is applied after every scattering, the weight of a photon package must be multiplied by a new biasing factor every time. The result is that the distribution of package weights becomes wider, which increases the noise. In those cases, other techniques such as diffusion approximation might be more appropriate \citep{1984JCoPh..54..508F, 2009A&A...497..155M, 2010A&A...520A..70R}.

One way to quantify the value of biasing and to investigate the circumstances in which potential drawbacks become relevant, is to apply it to a range of radiative transfer problems with a sufficiently large variety in geometry and optical depth. Such a suite of simulations is currently ongoing: TRUST\footnote{\url{http://ipag.osug.fr/RT13/RTTRUST/index.php}} \citep[Transport of Radiation through a dUSTy medium:][]{2015A&A...580A..87C, Gordon2016} is a suite of 3D dust radiative transfer benchmark simulations that is designed to cover the various numerical problems arising in dust radiative transfer, to further develop and improve the existing codes, and to test newly developed codes. In fact, the first of the TRUST benchmark tests \citep{Gordon2016} was the direct motivation for the novel biasing experiments presented here. In the course of the larger suite of TRUST benchmark simulations, the recipes presented here will be applied and evaluated more extensively than possible in this paper.

\begin{acknowledgements}
M.B. and P.C. acknowledge the financial support from CHARM (Contemporary physical challenges in Heliospheric and AstRophysical Models), a Phase-VII Interuniversity Attraction Pole program organised by BELSPO, the BELgian federal Science Policy Office.  T.L. acknowledges the support from the Swedish National Space Board (SNSB). M.J. acknowledges the support of the Academy of Finland Grant no. 285769. R.K. acknowledges financial support by the Emmy-Noether-Program of the German Research Foundation (DFG) under grant no. KU 2849/3-1. This work was supported by DustPedia, a collaborative focused research project supported by the European Union under the Seventh Framework Programme (2007-2013) call (proposal No. 606824). The participating institutions are: Cardiff University, UK; National Observatory of Athens, Greece; Ghent University, Belgium; Universit\'e Paris Sud, France; National Institute for Astrophysics, Italy; and CEA (Paris), France. 
\end{acknowledgements}

\bibliographystyle{aa} 
\bibliography{References}

\end{document}